# Hot Carrier and Surface Recombination Dynamics in Layered InSe Crystals

Chengmei Zhong,[1,2] Vinod K. Sangwan,[1] Joohoon Kang,[1] Jan Luxa,[3] Zdeněk Sofer,[3] Mark C. Hersam,[1,2,*] and Emily A. Weiss[1,2,*]

1. *Department of Materials Science and Engineering, Northwestern University, Evanston, IL 60208*

2. *Department of Chemistry, Northwestern University, Evanston, IL 60208*

3. *Department of Inorganic Chemistry, University of Chemistry and Technology Prague*
*Prague 6, Czech Republic*

*Corresponding authors: e-weiss@northwestern.edu; m-hersam@northwestern.edu

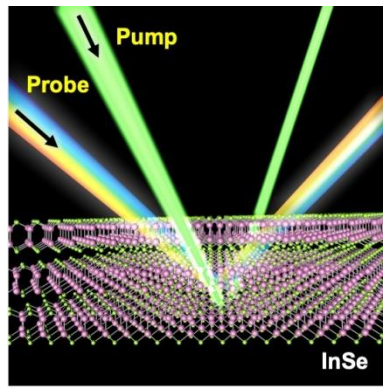

**Abstract:** Layered indium selenide (InSe) is a van der Waals solid that has emerged as a promising material for high-performance ultrathin solar cells. The optoelectronic parameters that are critical to photoconversion efficiencies, such as hot carrier lifetime and surface recombination velocity, are however largely unexplored in InSe. Here, these key photophysical properties of layered InSe are measured with femtosecond transient reflection spectroscopy. The hot carrier cooling process is found to occur through phonon scattering. The surface recombination velocity and ambipolar diffusion coefficient are extracted from fits to the pump energy-dependent transient reflection kinetics using a free carrier diffusion model. The extracted surface recombination velocity is approximately an order of magnitude larger than that for methylammonium lead-iodide perovskites, suggesting that surface recombination is a principal source of photocarrier loss in InSe. The extracted ambipolar diffusion coefficient is consistent with previously reported values of InSe carrier mobility.

Layered indium selenide (InSe) has been aggressively investigated for high-performance electronic and optoelectronics devices due to its high electron mobility and direct semiconducting band gap at all thicknesses greater than three layers.[1-15] Since the band gap of multilayer InSe is well-matched to the optimal Shockley-Queisser band gap, it is particularly promising for photovoltaic applications. InSe also possesses low thermal conductivity (due to weak electron-phonon coupling)[16], which, in lead-iodide perovskites,[17] leads to long hot carrier lifetimes (~100 ps) for a portion of the population; this slow thermalization could enable hot carriers to contribute to the photocurrent and photovoltage within an InSe-based solar cell.[18] Hot carrier solar cells are theoretically predicted to have open circuit voltages larger than the band gaps of their component semiconductors,[19-20] and, in principle, can operate above the Shockley-Queisser limit.[21] Layered bulk InSe was discovered over half a century ago[22] and was subsequently studied as an absorbing layer in Schottky solar cells with power conversion efficiencies exceeding 10%.[23-26] As a van der Waals solid, InSe has recently attracted renewed interest since exfoliation methods have shown that ultrathin flakes of InSe are promising two-dimensional (2D) semiconductors for application in ultrathin solar cells.[27-29] The fabrication of high-performance photovoltaic junctions with ultrathin active layers, however, requires minimization of non-radiative surface recombination, which, in turn, requires fundamental studies of the surface recombination velocity (SRV).[6]

Here, we report the ultrafast hot carrier dynamics and surface recombination in layered InSe crystals using broadband femtosecond transient reflection (TR) spectroscopy, as a function of excited carrier density and excitation energy. These measurements yield three key findings: i) the dependence of hot carrier dynamics in layered InSe crystals on the density of excited carriers is similar to that in lead-iodide perovskite materials, in which phonon scattering mediates cooling; ii) the ambipolar diffusion coefficient is 15.1 cm$^2$s$^{-1}$, consistent with previously reported values of InSe carrier mobility; and (iii) the SRV is $4.1\times10^4$ cm s$^{-1}$, smaller than that for unpassivated Si or GaAs, but larger than that for methylammonium lead-iodide perovskites.[30-31] We expect that the conclusions we present here for bulk layered InSe will be useful in characterizing carrier diffusion and surface recombination in ultrathin flakes of InSe, where surface-mediated processes have even more critical influence on device performance. With, for instance, transient absorption microscopy[32-33] the thickness dependence of these critical parameters could be extended to the 2D limit.

Layered InSe crystals were grown by the Bridgman method; the details are described in the Methods section (see Supporting Figures S1 and S2 for materials characterization).[7] TR spectroscopy was conducted on the InSe crystals, which were held in N$_2$ filled quartz cuvette to avoid ambient degradation (see Methods). In general, for direct band gap semiconductors similar to InSe, the refractive index is a factor of 5 to 20 larger than the extinction coefficient at most visible wavelengths.[34] Thus, unlike the more commonly acquired transient absorption (TA) spectrum, the intensities of features within the TR spectrum of bulk semiconductors are approximately proportional to the pump photon-induced changes in refractive index, rather than the pump photon-induced changes in extinction coefficient,[30-31, 34] and the positve and negative features in TR spectra cannot be simply interpreted as photoinduced absorptions and ground state bleaches, respectively. Figure 1A (black trace) shows the TR spectrum of a single crystal of InSe at 1 ps after pumping at 1.38 eV (bandgap ~1.25 eV). To correctly assign the features in the TR spectrum, we convert it to the corresponding TA spectrum using an inverse Hilbert transformation (iHT, or Kramers-Kronig transformation [30]), Figure 1A (red trace). The validity of the iHT process for obtaining TA data is confirmed by comparing the iHT of the TR spectrum to a directly acquired

TA spectrum of thin InSe flakes (still bulk-like) exfoliated from the same batch of the crystal (see Supporting Figure S3).

A comparison of the derived TA spectrum with the steady-state absorption spectra (Figure 1A, dashed blue trace) of exfoliated InSe flakes indicates that the narrow negative bleach feature in the TA spectrum at 2.42 eV, and the corresponding derivative-like feature in the TR spectrum, are attributable to the B exciton peak of InSe, which is the transition from a deep valence band state (VB2) to the conduction band-edge,[1] as shown in Figure 1B. Discrete and sharp peak features in TA/TR spectra usually indicate exciton formation (free carrier absorption features are usually broad and continuous); however, in the case of InSe, for which the exciton binding energy is only ~0.5 $k_B T$ at room temperature,[35] these sharp features are instead caused by the state-filling effect of free carriers. Specifically, existing free carriers force newly photoexcited carriers to occupy increasingly higher energy states through Pauli exclusion. This state-filling creates hypsochromic shifts in the absorption/reflection spectra, which then translate to first order derivative-like features in TR spectra.[36-37]

Figures 2A,C show that, when InSe is pumped at 1.38 eV (near the bandgap) such that minimal hot carriers are created, the shape of the TR spectrum is constant over the entire time window of the TR measurement (3 ns). Figures 2B,D show that, after pumping the sample at 3.06 eV (~1.9 eV above the band gap), which generates significant hot carrier population, the relative intensities of the positive and negative peaks in TR spectra evolve over time, on two distinct timescales (Supporting Figure S4 shows the normalized TR spectra). Before 5 ps (Figures 2B,D), the negative peak of the TR feature is significantly suppressed relative to the positive peak due to the additional TR contribution from hot carrier states. At a delay of ~5 ps, the excitonic features appear similar to those in the spectrum of InSe pumped at 1.38 eV, indicating completion of the cooling process.

The dynamics of thermalization are most easily followed at probe photon energies ($E_{probe}$) near 2.5 eV, where the hot carrier contribution to the TR spectrum is maximized. Figure 3A shows kinetic traces for hot carrier cooling, isolated by subtracting the early timescale (< 6 ps) TR kinetic trace obtained with $E_{pump}$ = 1.38 eV from the TR kinetic traces obtained with $E_{pump}$ = 3.06 eV, 2.06 eV, or 1.77 eV at the same excited carrier density (see the raw TR kinetics at $E_{probe}$ = 2.5 eV in Supporting Figure S6A). For an excited carrier density of $N_{pump} = 5 \times 10^{18}$ cm$^{-3}$, we measure hot carrier cooling times of 0.61 – 0.99 ps (see the legend of Figure 3A), and no discernible dependence of the cooling time on the energy of the pump photons.

Figures 3B,C show the dependence of the relaxation dynamics of hot carriers created with $E_{pump}$ = 3.06 eV as a function of $N_{pump}$ (see the raw TR kinetics at $E_{probe}$ = 2.5 eV in Supporting Figure S6B). The lifetime for hot carrier cooling increases with increasing $N_{pump}$ above a threshold density of ~$10^{19}$ cm$^{-3}$, consistent with a hot-phonon bottleneck model. In this model, a higher hot carrier density results in an increased population of non-equilibrium phonons, which increases the chances of phonon reabsorption by thermalized free carriers and consequently reduces the net carrier thermalization rate.[36] The cooling times of hot electrons in InSe single crystals extracted from these data (~1 – 3.5 ps) are comparable to the reported values in methylammonium lead-iodide perovskites under similar excitation densities.[32, 36] and significantly slower than those for bulk Si[38] or GaAs [39]. This result is reasonable since the large mass ratio of anion-to-cation in both of these materials (~1.4:1 for In$^{2+}$:Se$^{2-}$ and ~2:1 for Pb$^{2+}$:I$^-$)[36] results in large phononic bandgaps that slow hot carrier cooling.[40]

At later times (*e.g.*, 100 ps), the overall TR signal for the B exciton is lower and some of the asymmetry of the features in the TR spectrum returns (Figure 2B and Supporting Figure S5), but we suspect these changes are due to longer-timescale surface recombination processes discussed directly below.

Figures 4A,B show the free carriers dynamics in single crystal InSe at later times (between 6 ps and 3 ns), when thermalization of carriers is complete. An $E_{probe}$ of 2.38 eV and $E_{pump}$ of 1.77 eV were chosen because, with these parameters, the TR spectra have maximum signal-to-noise ratio and the TR intensity has a linear dependence on $N_{pump}$ (see Supporting Figure S7). According to a state-filling model,[30, 37] such a linear dependence is a strong indication that the TR signal is predominantly coming from free carriers, which are either free electrons in the lowest conduction band level (CB) or free holes in a deep valence band level (VB2) shown in Figure 1B. The longer-time (>6 ps) dynamics of recombination for free carriers in InSe are faster with higher $E_{pump}$ (Figure 4A), but the dependence of these dynamics on $N_{pump}$ is negligible (Figure 4B). The fact that free carrier dynamics have qualitatively different $E_{pump}$ and $N_{pump}$ dependencies at early times (<6 ps) and later times (>6 ps) implies that they are governed by different mechanisms.

The free carrier dynamics of InSe after thermalization can be explained by a free carrier diffusion model that includes surface recombination terms proposed by Yang *et al.*[30-31, 41] Like other bulk semiconductors,[30-31] the absorption coefficient of an InSe single crystal monotonically increases with photon energy below 3.3 eV.[22] Therefore, a pump pulse with higher $E_{pump}$ leads to a larger spatial gradient of the initial free carrier density near the sample surface, which produces a larger carrier diffusion rate and faster surface recombination dynamics. The dynamics of the TR signal are therefore sensitive to the pump energy because the TR measurement exclusively probes carriers within ~20 nm of the semiconductor surface ($\lambda_{probe}/4\pi n$, where $\lambda_{probe} \equiv$ probe wavelength, $n \equiv$ refractive index[30]). We do not observe a dependence of free carrier dynamics on $N_{pump}$ because density-dependent bulk recombination of free carriers in inorganic semiconductors usually occurs on timescales much larger than our measurement time window (>100 ns).[30, 42]

Within the Yang model described above, the ambipolar diffusion coefficient $D$ (i.e., the average of electron and hole diffusion constants) and surface recombination velocities (SRVs)[30-31, 41] can be extracted by fitting the TR kinetic traces at $E_{probe}$ = 2.38 eV between 6 ps and 3 ns with equation 1:

$$\frac{\Delta R}{R}(t) = \frac{1}{SRV - \alpha D}[-\alpha D w(\alpha\sqrt{Dt}) + SRV w(SRV\sqrt{\frac{t}{D}})] \quad (1)$$

In equation 1, $t$ is the pump-probe time delay, $w(x) = e^{x^2} erfc(x)$, and $\alpha$ is the absorption coefficient of InSe at the corresponding pump photon energy: $\alpha$ = $5\times10^4$/$5\times10^3$/$1\times10^3$ cm$^{-1}$ at 3.06/2.04/1.38 eV pump energies, respectively.[22] We fit all of the kinetic traces in Figure 4A simultaneously with equation 1 using a nonlinear least-squares global fitting routine where the values of $SRV$ and $D$ are shared among all traces. We obtain an $SRV$ value of $4.1\times10^4$ cm s$^{-1}$, which is approximately a factor of ten higher than that for methylammonium lead-iodide perovskite materials,[31] but a factor of 10 – 100 lower than those for inorganic bulk semiconductors used in high efficiency solar cells (e.g., $SRV = 2.4\times10^5$ cm s$^{-1}$ for unpassivated p-type Si [43], $SRV = 1.2\times10^6$ cm s$^{-1}$ for n-type Si [43], $SRV = 8.5\times10^5$ cm s$^{-1}$ for GaAs [44]). This observation suggests that surface recombination in InSe is at an intermediate level among common inorganic solar cell materials, and partly explains why InSe solar cells only have moderate power conversion efficiencies despite their excellent optical and electronic properties.[23-24] Since the efficiency of Si and GaAs solar cells improve greatly after surface passivation,[45-46] it is reasonable to expect that similar improvements can be achieved for InSe-based solar cells once a reliable surface passivation scheme is developed.

We extract a $D$ value of 15.1 cm$^2$s$^{-1}$, which is also high compared to methylammonium lead-iodide perovskites due to the higher carrier mobility of InSe. Using the Einstein relation – $\mu$ =

$eD/k_BT$, where, $e$, $k_B$, and $T$ are the electronic charge, the Boltzmann constant, and temperature, respectively – we estimate the average photocarrier mobility ($\mu$) near the surface of InSe to be ~500 cm$^2$V$^{-1}$s$^{-1}$. This value is similar to the highest reported values of intralayer electron and hole mobilities at room temperature in single crystal InSe (~1000 cm$^2$V$^{-1}$s$^{-1}$ for electrons,[1, 3, 5] ~30 cm$^2$V$^{-1}$s$^{-1}$ for holes[47]). The low mobility value we measure (relative to published values) is possibly attributable to the fact that the TR measurement is performed at an oblique reflection angle, which causes our estimate to be the average of interlayer mobility ($10^{-1}$~$10^0$ cm$^2$V s$^{-1}$)[48] and intralayer mobility (~$10^3$ cm$^2$V s$^{-1}$) .

In conclusion, we have studied the hot carrier and surface recombination dynamics of InSe single crystals with femtosecond transient reflection (TR) spectroscopy. The hot carrier lifetime is dependent on excited carrier density, indicating that the cooling process is caused by phonon scattering. The relatively long hot carrier lifetime (~1 ps for a carrier density of $10^{18}$ cm$^{-3}$) is likely caused by the large phononic band gap created by the large mass ratio between anion (In$^{2+}$) and cation (Se$^{2-}$). We further determined the values of critical parameters for solar cell device engineering, specifically the ambipolar diffusion coefficient ($D$) and the surface recombination velocity ($SRV$), by modeling the dynamics of free carriers after they have thermalized (>6 ps) using drift-diffusion equations that include surface recombination.[30-31] The measured ambipolar diffusion constant $D$ is 15.1 cm$^2$s$^{-1}$, which is consistent with published values of carrier mobility in InSe. Our extracted $SRV$ value of 4.1×10$^4$ cm s$^{-1}$ is at an intermediate level among common inorganic solar cell materials, and suggests that the development of surface passivation schemes for InSe could minimize surface recombination and improve power conversion efficiencies. An additional question is whether there is any impact on the observed hot carrier dynamics due to depopulation of VB2 into VB1. In principle, it is possible to combine the TR spectra and kinetics of the A-exciton regions in the near-infrared with the visible-region TR data to decouple the VB2-to-VB1 transition from hot carrier dynamics. In practice, however, the much lower signal-to-noise and time-dependent peak shifts in the NIR data prevent us from doing so here; some further experimental optimization is warranted to pursue this issue. Overall, this work reveals the key photophysical properties of InSe, which will inform ongoing efforts to realize high-performance optoelectronic and photovoltaic applications based on this emerging van der Waals solid.

**METHODS**

**InSe Single Crystal Growth and TR Sample Preparation.** InSe crystal growth is described elsewhere.[7] Detailed structural characterization of InSe single crystals are given in the Supporting Information (Figures S1 and S2). For TR measurements, small pieces of atomically flat InSe crystals with surface areas >0.1 cm$^2$ and thicknesses >0.1 mm were cleaved perpendicular to the surface normal from the InSe crystal ingot by a pair of tweezers inside a nitrogen glovebox.. The small pieces were then taped onto glass substrates and deposited into quartz cuvettes, which were sealed with Teflon tape and vacuum grease inside the glovebox before being taken into ambient conditions for TR measurements.

**Transient Reflection Spectroscopy Measurements in the Visible Wavelength Region.** Visible transient reflection spectroscopy measurements were performed following previously published protocols[49-50] with the addition of two silver mirrors before the probe detector to adjust the beam path for reflection geometries. The inverse Hilbert transform of TR spectral data was performed with the commercial module incorporated in OriginPro 8.6 (OriginLab).

**FIGURES**

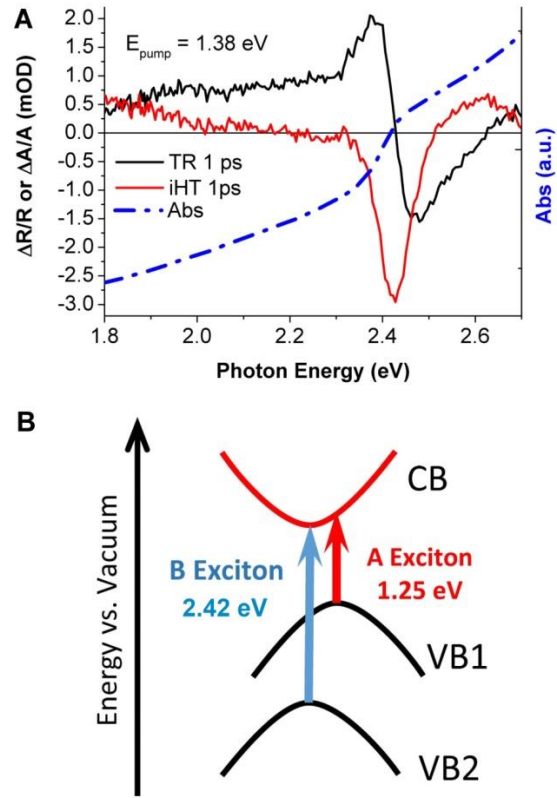

**Figure 1.** **(A)** TR spectrum of an InSe single crystal pumped at 1.38 eV photon energy (black), and the inverse Hilbert transform (iHT) of the TR spectrum (red), which represents the surface transient absorption spectrum through the Kramers–Kronig relationship. The dashed blue trace is the ground state absorption spectrum of an exfoliated thin film taken from the same batch of InSe crystals. **(B)** Schematic of electronic transitions in single crystal InSe corresponding to A and B exciton absorption peaks according to the theoretical calculations in Ref [1]. CB stands for the conduction band, VB1 and VB2 stand for the highest and the second highest valance bands, respectively. The derivative-like feature centered around 2.4 eV in the TR spectrum in **(A)** corresponds to a ground state bleach in the TA spectrum, with an energy that matches the transition from VB2 to CB in **(B)**. The broad positive feature from 1.8 eV to 2.2 eV in the TR spectrum in **(A)** is a photoinduced absorption feature (probably due to free carrier absorptions in InSe) in the TA spectrum.

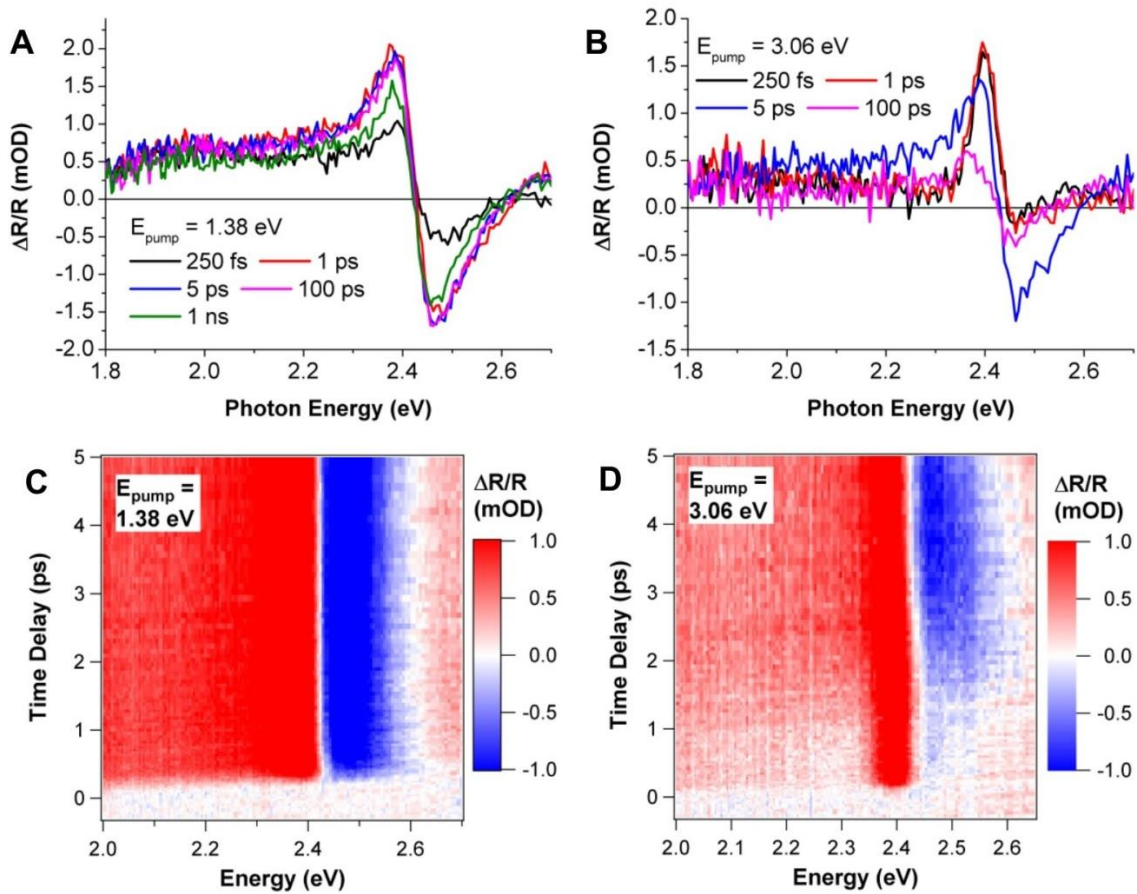

**Figure 2. (A, B)** TR spectra of an InSe single crystal in the visible region, at time delays of 250 fs, 1 ps, 5 ps, and 100 ps after excitation at 1.38 eV pump energy (near the bandgap, **A**) and 3.06 eV pump energy (1.9 eV above the bandgap, **B**), respectively. **(C, D)** Pseudo-color 2D images of TR spectra of an InSe single crystal pumped at 1.38 eV (**C**) or 3.06 eV (**D**). The horizontal axis is the probe photon energy (eV), and the vertical axis is the pump-probe time delay (ps). The color intensity indicated by the scale bar represents the TR intensity (mOD).

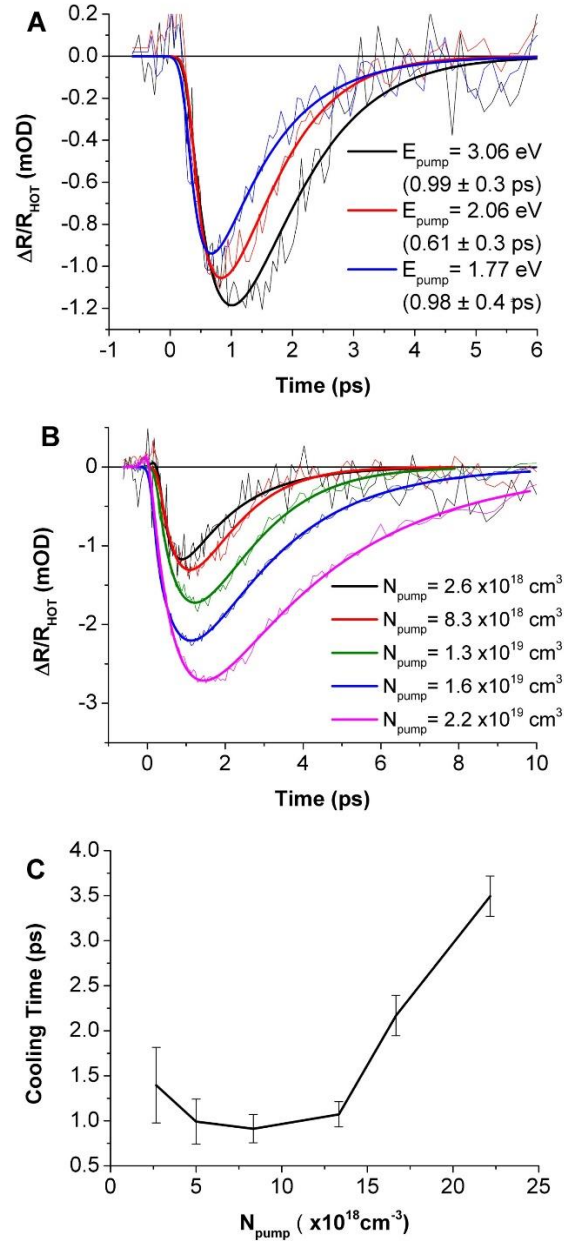

**Figure 3. (A)** Hot carrier thermalization dynamics of single crystal InSe extracted from Figure S4A by subtracting the raw TR kinetic trace acquired at $E_{pump} = 1.38$ eV from the kinetic traces acquired at $E_{pump} = 3.06$ eV, 2.06 eV, or 1.77 eV, as indicated. The cooling times from exponential fits of the kinetic traces are shown in the legend. $N_{pump}$ is fixed at $5\times10^{18}$ cm$^{-3}$ for all pump energies. **(B)** Hot carrier thermalization dynamics of single crystal InSe extracted from Figure S4B for $E_{pump} = 3.06$ eV, with $N_{pump} = 2.6, 5, 8.3, 13, 16,$ or $22 \times 10^{18}$ cm$^{-3}$, respectively. The corresponding thermalization lifetimes are plotted vs. $N_{pump}$ in **(C)**.

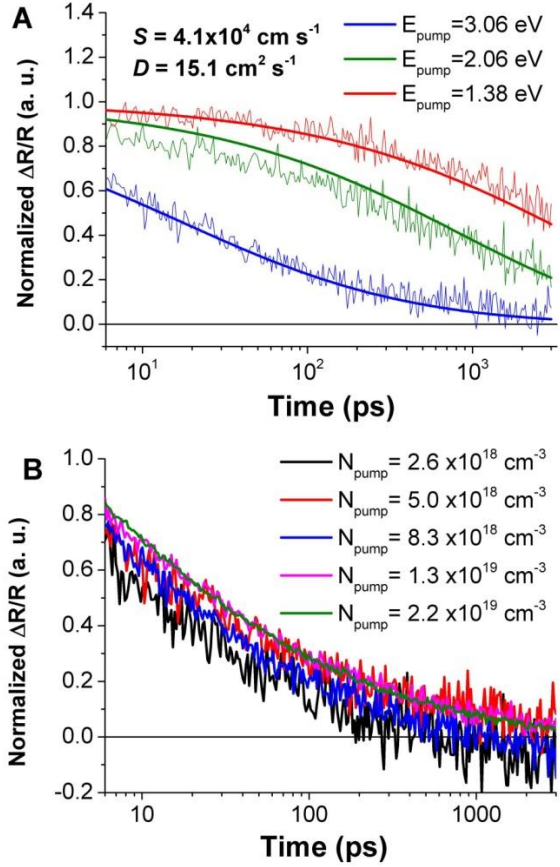

**Figure 4.** **(A)** $E_{pump}$ dependent TR kinetic traces (normalized at their peak values at ~2 ps) from 6 ps to 3 ns of an InSe single crystal, pumped at 2.38 eV from the TR spectrum of the same InSe crystal, and pumped with a fixed $N_{pump}$ of $5\times10^{18}$ cm$^{-3}$; the nonlinear least-squares global best-fit curves from 6 ps to 3 ns using equation 1 and the extracted values of SRV (**S**) and diffusion constant (**D**), are shown. **(B)** $N_{pump}$ dependent normalized TR kinetic traces (normalized at a maximum at ~2 ps) from 6 ps to 3 ns extracted at $E_{probe} = 2.38$ eV probe energy and $E_{pump} = 3.06$ eV.

**SUPPORTING INFORMATION**

TEM and XRD characterizations of InSe single crystal samples; additional transient reflection spectroscopy data and analysis; and Figures S1-S7.

**ACKNOWLEDGMENTS**

This research was supported by the Materials Research Science and Engineering Center (MRSEC) of Northwestern University (NSF DMR-1720139) (spectroscopy) and the Center for Light Energy Activated Redox Processes (LEAP), an Energy Frontier Research Center funded by the U. S. Department of Energy, Office of Science, Basic Energy Sciences under Award No. DE-SC0001059 (sample preparation and characterization). Z. S. and J. L. were supported by the Czech Science Foundation (GACR No. 17-11456S), Neuron Foundation for scientific support and by the project Advanced Functional Nanorobots (reg. No. CZ.02.1.01/0.0/0.0/15_003/0000444 financed by the EFRR).